\documentclass[prd,nofootinbib,showpacs,preprint,aps]{revtex4-1}

\usepackage{amsfonts}
\usepackage{mathrsfs}
\usepackage{epsfig}
\usepackage{graphicx}%
\usepackage{dcolumn}
\usepackage{amsmath}
\usepackage{color}
\usepackage{color}
\newcommand\A{\%}
\makeatletter
\def\btt#1{\texttt{\@backslashchar#1}}%
\DeclareRobustCommand\bblash{\btt{\@backslashchar}}%
\makeatother
\usepackage{subfigure}
\usepackage{amsmath}
\usepackage{amsfonts}
\usepackage{amssymb}

\usepackage[applemac]{inputenc}
\usepackage[T1]{fontenc}
\usepackage{amssymb}
\usepackage{amsfonts}
\usepackage{amsfonts}
\usepackage{graphicx}

\usepackage{bbm}

\begin{document}
\title{How to accomplish inflation and a constant dark energy density}
\author{Changjun Gao}
\email{gaocj@nao.cas.cn}
\affiliation{National Astronomical Observatories, Chinese Academy of Sciences, 20A Datun Road, Beijing 100101, China}

\begin{abstract}\baselineskip=18pt
It is commonly assumed that the Lagrangian of multi-field theories of gravity contains the sum of kinetic terms of scalar fields. However, we propose here the Lagrangian contains not the sum but the quotient of kinetic terms. With this novel change, we find the inflationary universe can be acquired with two scalars without the slow-roll approximation. On the other hand, a constant dark energy density can also be achieved with two scalars while without the need of Einstein cosmological constant.

\vskip 1in
\leftline{\emph{Essay written for the Gravity Research Foundation 2022 Awards for Essays on Gravitation.}}
\leftline{\emph{Submitted on Mar. 10, 2022}}
\leftline{\emph{gaocj@nao.cas.cn}}
\end{abstract}

\maketitle

\section{introduction}
It is well-known that inflation models provide an elegant solution to the horizon and flatness problems \cite{guth:1981,linde:1982} as well as a mechanism to seed quantum
fluctuations which is in excellent agreement with the latest observations \cite{ak:2018}. Inflation therefore becomes the standard paradigm for describing the physics of the very early universe. But the nature of the inflaton fields remains an open question. On the other hand, seeking for the physical mechanism for the late-time
cosmic acceleration \cite{perl:1998,adam:1998} is one of the most exciting
challenges in cosmology. The Einstein cosmological constant is phenomenologically referred to as the
simplest explanation \cite{fri:2008,bart:2010}. But due to the fine-tuning problem of the cosmological
constant \cite{jer:2012,bur:2015}, there remains an extensive effort dedicated to
the study of possible alternatives (see Refs. \cite{cop:2006,phi:2016,tim:2012,Nojiri:2010wj,Bamba:2012cp,Nojiri:2017ncd,aus:2015} for reviews).

In this paper we report a novel Lagrangian which is composed of the quotient of kinetic terms and scalar fields, which we call quotient scalar fields. With this novel Lagrangian, the inflationary universe can be naturally acquired with two scalars without the slow-roll approximation which is assumed in single inflaton field. On the other hand, a constant dark energy density can also be achieved with two scalars while without the need of Einstein cosmological constant. Actually, there is a reason why a single scalar field model does not work as dark energy. It is found \cite{ban:2021} that replacing cosmological constant with quintessence lowers the present-day Hubble constant while observations prefer a higher one. This motivates us to consider two scalar fields.

\section{cosmic evolution of single scalar field}
We start from the gravity theory with one scalar field $\phi$. The Lagrangian is
\begin{eqnarray}
\mathscr{L}=\sqrt{-g}\left[\frac{R}{16\pi}-X+V\left(\phi\right)\right]\;,
\end{eqnarray}
with $X=-\frac{1}{2}\nabla_{\mu}\phi\nabla^{\mu}\phi$ the kinetic energy of scalar field $\phi$. $R$ is the Ricci scalar and $V(\phi)$ is the scalar potential. In the background of Friedmann-Robertson-Walker (FRW) Universe which has the following line-element
\begin{eqnarray}
ds^2=-N\left(t\right)^2dt^2+a\left(t\right)^2\left(dr^2+r^2d\theta^2+r^2\sin^2\theta d\varphi^2\right)\;,
\end{eqnarray}
where $N(t)$ is the lapse function and $a(t)$ is the scale factor, the Lagrangian becomes
\begin{eqnarray}\label{lag}
\mathscr{L}=a^3\left[\frac{1}{16\pi}\left(-\frac{6\ddot{a}}{aN}+\frac{6\dot{N}\dot{a}}{aN^2}-\frac{6\dot{a}^2}{a^2N}\right)-\frac{1}{2N}\dot{\phi}^2+NV\left(\phi\right)\right]\;.
\end{eqnarray}
Here the dot denotes the derivative with respect to cosmic time $t$. Variation of the corresponding action $S=\int \mathscr{L}dt$ with respect to $N$, $a$ and $\phi$ give the Friedmann equation, the acceleration equation and the scalar field equation, respectively, (finally setting $N=1$)
\begin{eqnarray}
3H^2=8\pi\left(\frac{1}{2}\dot{\phi}^2+V\right)\;,\ \ \ \ 2\dot{H}+3H^2=-8\pi\left(\frac{1}{2}\dot{\phi}^2-V\right)\;,\ \ \ \ddot{\phi}+3H\dot{\phi}+V_{,\phi}=0\;,
\end{eqnarray}
where $``,"$ denotes the derivative with respect to the scalar field $\phi$ and $H=\dot{a}/a$ is the Hubble parameter. From the Friedmann equation and the acceleration equation, we read out the energy density and pressure of scalar field
\begin{eqnarray}
\rho=\frac{1}{2}\dot{\phi}^2+V\;,\ \ \ \ \ p=\frac{1}{2}\dot{\phi}^2-V\;.
\end{eqnarray}
It is apparent the energy density is the sum of kinetic energy $\dot{\phi}^2/2$ and potential energy $V$. The inflation universe models usually require that the energy density is contributed by only potential energy \cite{bau:0907}. Then how to accomplish this? It is generally assumed that the scalar field rolls down the hill of scalar potential very slowly such that the kinetic energy can be safely neglected, i.e. $\dot{\phi}^2/2\ll V$. However, we shall propose the alternative of quotient scalar fields in the next section.
\section{inflation}
Observing the Lagrangian Eq.~(\ref{lag}), we find the kinetic energy term $-\frac{\dot{\phi}^2}{2N}$ has the factor of $1/N$ while the potential $V$ has the factor of $N$. If the kinetic energy term does not have the lapse function $N$, it would not make contribution to the energy density and the energy density is uniquely contributed by the scalar potential. Then how to eliminate the lapse function $N$ in the kinetic term? We propose the gravity theory with two scalars, $\phi$ and $\psi$, for example
\begin{eqnarray}\label{laginf1}
\mathscr{L}=\sqrt{-g}\left[\frac{R}{16\pi}+\sqrt{\frac{X^2}{Y}}+V\left(\phi\right)\right]\;,
\end{eqnarray}
with $Y=-\frac{1}{2}\nabla_{\mu}\psi\nabla^{\mu}\psi$ the kinetic energy of $\psi$. The novel property of the model is that the kinetic term is not the sum of kinetic terms $X$ and $Y$ but the square root for quotient of kinetic terms $X^2$ and $Y$. Therefore, we call them quotient scalar fields. Then the lapse function does not appear in the kinetic term because of this novel change. The corresponding energy momentum tensor is
\begin{eqnarray}
T_{\mu\nu}=\left(\sqrt{\frac{X^2}{Y}}+V\right)g_{\mu\nu}+\frac{\nabla_{\mu}\phi\nabla_{\nu}\phi}{\sqrt{Y}}-\frac{X\nabla_{\mu}\psi\nabla_{\nu}\psi}{2Y\sqrt{Y}}\;.
\end{eqnarray}
The equations of motion are
\begin{eqnarray}
3H^2=8\pi V\;,\ \ \ 2\dot{H}+3H^2=-8\pi\left(-\frac{\sqrt{2}}{2}\cdot\frac{\dot{\phi}^2}{\dot{\psi}}-V\right)\;,
\end{eqnarray}
\begin{eqnarray}
\left(\frac{a^3\dot{\phi}^2}{\dot{\psi}^2}\right)^{\cdot}=0\;,\ \ \ \ \ \ \ \
\frac{\sqrt{2}}{a^3}\left(\frac{a^3\dot{\phi}}{\dot{\psi}}\right)^{\cdot}-V_{,\phi}=0\;.
\end{eqnarray}
When $V\propto\phi^{\frac{4}{2-3n}}$, we find the scale factor evolves as $a\propto t^{n}$.
For large $n$, we will acquire an inflating universe. On the other hand, if we choose
$V=\frac{\Lambda}{\pi}\left(\textsl{erfi}^{-1}{\phi}\right)^2$, we will find the scale factor evolves as $a\propto e^{\Lambda t^2}$.
Here ${\textsl{erfi}}^{-1}{\phi}$ is the inverse imaginary error function of $\phi$ and $\Lambda$ is a positive constant. The inverse imaginary error function ${\textsl{erfi}}^{-1}{\phi}$ behaving like the quadratic function $y=x^2$, is non-negative and has a global minimum at $\phi=0$. It is straightforward to extend the Lagrangian Eq.~(\ref{laginf1}) into the other cases, such as
\begin{eqnarray}\label{laginf2}
\mathscr{L}=\sqrt{-g}\left[\frac{R}{16\pi}+\sqrt{\frac{X^{i}}{Y^{i-1}}}+V\left(\phi\;, \psi\right)\right]\;,
\end{eqnarray}
where $i\geq 2$ is positive integer. The cosmic energy density in these cases is uniquely contributed by the scalar potential $V$ without the assumption of slow-roll approximation.
\section{Perturbations}
It seems unusual that the Lagrangian Eq.~(\ref{laginf1}) has the quotient but not the sum of kinetic terms. Can we transform the quotient to the sum? The answer is yes.  Introducing two auxiliary scalar fields $\Phi$ and $\Psi$, we find the Lagrangian Eq.~(\ref{laginf1}) can be rewritten as
\begin{eqnarray}\label{cc3}
\mathscr{L}=\sqrt{-g}\left[\frac{R}{16\pi}+\Psi X+\Phi Y-\frac{\Phi}{\Psi^2}+V\left(\phi\right)\right]\;.
\end{eqnarray}
Then four scalars $\phi\;,\ \psi\;,\ \Phi$ and $\Psi$ are present and the quotient of kinetic terms disappears. We recognize immediately that it belongs to the generalized  multi-field theories \cite{langlois:2008} where the linear perturbations of inflation are already completed. We point out that there is a significant  difference from the generalized  multi-field theories \cite{langlois:2008}. There are no kinetic terms for the scalar field $\Phi$ and $\Psi$ and the two scalars are auxiliary fields. Using the results in Ref.~\cite{langlois:2008}, we find the square of effective speed of sound is $c_s^2=1$. The tensor-to-scalar ratio at the sound horizon crossing is $r=2\epsilon=-{2\dot{H_{\ast}}}/{H_{\ast}^2}={{2}/{n}}=2/N_{\ast}$ for the scalar potential $V\propto\phi^{\frac{4}{2-3n}}$.  Here the subscript ``$\ast$" denotes the time of horizon crossing and $N_{\ast}$ is the e-folding number at the horizon crossing. On the other hand, we find the tensor-to-scalar ratio $r=2\epsilon=-\frac{1}{N_{\ast}}$ for the scalar potential $V=\frac{\Lambda}{\pi}\left(\textsl{erfi}^{-1}{\phi}\right)^2$. For
most of the inflation models, for example, the chaotic inflation \cite{linde:1983}, we have $50<N_{\ast}<60$ \cite{akr:2020}.
Planck 2018 \cite{akr:2020} together with BICEP2/Keck Array BK15 \cite{ade:2018} data require $r<0.06$ at $95{\A}$ CL. Substituting the e-folding numbers $N_{\ast}$ into the expression of tensor-to-scalar ratio $r$ shows the two inflationary models are consistent with the observations very well.
\section{a constant dark energy density in FRW universe}
In this section, we show how to achieve a constant dark energy density with two scalars. To this end, we consider the gravity theory with quotient scalar fields, $\phi$ and $\psi$ as follows
\begin{eqnarray}\label{cc1}
\mathscr{L}=\sqrt{-g}\left[\frac{R}{16\pi}+{\frac{X}{Y}}+V\left(\phi\;, \psi\right)+\rho_m\right]\;.
\end{eqnarray}
Now both the kinetic term $X/Y$ and the potential term $V$ have no lapse function $N$ and the scale factor $a$ in the background of FRW Universe. Therefore they make contributions to the energy density and pressure in the same manner. The energy momentum tensor is
\begin{eqnarray}\label{emt}
T_{\mu\nu}=\left(\frac{X}{Y}+V\right)g_{\mu\nu}+\frac{\nabla_{\mu}\phi\nabla_{\nu}\phi}{Y}-\frac{X\nabla_{\mu}\psi\nabla_{\nu}\psi}{Y^2}\;.
\end{eqnarray}
The equations of motion are
\begin{eqnarray}
3H^2=8\pi \left(\frac{\dot{\phi}^2}{\dot{\psi}^2}+V+\rho_m\right)\;,\ \ \ \
2\dot{H}+3H^2=8\pi \left(\frac{\dot{\phi}^2}{\dot{\psi}^2}+V-p_{m}\right)\;,
\end{eqnarray}
\begin{eqnarray}
\left(\frac{2a^3\dot{\phi}}{\dot{\psi}^2}\right)^{\cdot}-a^3V_{,\phi}=0\;,\ \ \ \
\left(\frac{2a^3\dot{\phi}^2}{\dot{\psi}^3}\right)^{\cdot}+a^3V_{,\psi}=0\;,\ \ \
\frac{d\rho_m}{dt}+3H\left(\rho_m+p_m\right)=0\;.
\end{eqnarray}
Given the equation of state for matter $\omega_m=p_m/\rho_m$ and the scalar potential $V$, the equations of motion are closed. We find the energy density contributed by the two scalars is exactly a constant
\begin{eqnarray}
\rho_X=\frac{\dot{\phi}^2}{\dot{\psi}^2}+V=\Lambda\;,
\end{eqnarray}
although the scalar fields $\phi$ and $\psi$ are dynamic. For example, we have $\dot{\phi}\propto a^3$ and $\dot{\psi}\propto a^3$  in the absence of the scalar potential $V$. It is straightforward to extend the Lagrangian Eq.(\ref{cc1}) to the case of multiple scalars, for example three scalars $\phi, \psi, \varphi$
\begin{eqnarray}\label{cc2}
\mathscr{L}=\sqrt{-g}\left[\frac{R}{16\pi}+F\left(\phi\;, \psi\;, \varphi\;, \chi_1\;, \chi_2\right)+\rho_m\right]\;,
\end{eqnarray}
with $\chi_1$ and $\chi_2$ defined by
\begin{eqnarray}
\chi_1=\frac{X}{Y}\;, \ \ \ \chi_2=\frac{Z}{Y}\;,\ \ \  Z=-\frac{1}{2}\nabla_{\mu}\varphi\nabla^{\mu}\varphi\;.
\end{eqnarray}
It is found that these scalars contribute a constant dark energy density and pressure in the background of FRW universe. In fact, even if in the anisotropic but homogeneous universe such that
\begin{eqnarray}
ds^2=-dt^2+a_1\left(t\right)^2dx^2+a_2\left(t\right)^2dy^2+a_3\left(t\right)^2dz^2\;,
\end{eqnarray}
with $a_1$, $a_2$ and $a_3$ different scale factors, these scalar fields still  contribute a constant dark energy density and pressure.

\section{a constant dark energy density in black hole spacetime}
In this section, we show the gravity theory Eq.~(\ref{cc1}) gives a constant dark energy density in black hole spacetime. To this end, we consider the static and spherically symmetric spacetime with the metric
\begin{eqnarray}
ds^2=-U\left(r\right)dt^2+\frac{f\left(r\right)^2}{U\left(r\right)}dr^2+r^2d\Omega_2^2\;.
\end{eqnarray}
We obtain three Einstein equations from Lagrangian Eq.~(\ref{cc1})
\begin{eqnarray}
f^{'}=0\;,
\end{eqnarray}
\begin{eqnarray}
2U^{'}f\psi^{'2}-2Uf^{'}\psi^{'2}-rU^{'}f^{'}\psi^{'2}+rf\psi^{'2}U^{''}+rf^3\phi^{'2}+rVf^3\psi^{'2}=0\;,
\end{eqnarray}
\begin{eqnarray}
2U\psi^{'2}+2rU^{'}\psi^{'2}-2f^2\psi^{'2}+r^2f^2\phi^{'2}+r^2Vf^2\psi^{'2}=0\;,
\end{eqnarray}
and the equations of motion for $\phi$ and $\psi$
\begin{eqnarray}
2rf^{'}\psi^{'}\phi^{'}+4f\psi^{'}\phi^{'}+2 rf\psi^{'}\phi^{''}-4 rf\phi^{'}\psi^{''}-rf\psi^{'3}V_{,\phi}=0\;,
\end{eqnarray}
\begin{eqnarray}
4f\psi^{'}\phi^{'2}+2 rf^{'}\psi^{'}\phi^{'2}+4 r\psi^{'}\phi^{'}\phi^{''}-6 rf\psi^{''}\phi^{'2}+rf\psi^{'4}V_{,\psi}=0\;.
\end{eqnarray}
From the three Einstein equations we obtain
\begin{eqnarray}
f=1\;,\ \ \ \ U=1-\frac{2M}{r}-\frac{1}{3}\Lambda r^2\;,
\end{eqnarray}
where $M$ and $\Lambda$ are integration constants which have the physical meaning of mass and cosmological constant.
It is exactly the Schwarzschild-de Sitter solution. In the absence of scalar potential $V$, the scalar fields $\phi$ and $\psi$ are
\begin{eqnarray}
\phi=r^3\;,\ \ \ \psi=\sqrt{\frac{1}{2\Lambda}}r^3\;.
\end{eqnarray}
Therefore, the Schwarzschild-de Sitter spacetime is more than the solution of Einstein cosmological constant, it is the solution of quotient scalar fields.

\section{conclusion and discussion}

It is amazing that once the Lagrangian of multi-field theory is endowed with the quotient of kinetic terms. On the one hand, the inflationary universe can be readily accomplished without slow-roll approximation. On the other hand, a constant dark energy density can be accomplished and the cosmological constant is substituted.
Are there similar forms of Lagrangian in classical physics? The answer is yes.
We remember that the Lagrangian for massive particles is $\mathscr{L}=\sqrt{1-v^2/c^2}$ in Special Relativity. $v$ and $c$ are the velocities of massive particles and photons, respectively. Inspired by this Lagrangian, we can consider $\mathscr{L}=\sqrt{-g}[K(\phi\;,\psi)\sqrt{1-X/Y}+V(\phi\;,\psi)]$. $X$ and $Y$ are understood as the square of velocities for scalar particles $\phi$ and $\psi$. When $Y$ is a constant and $\psi$ vanishing, it reduces to the generalized tachyon field Lagrangian. In the background of FRW universe, we have $\mathscr{L}=a^3N[K(\phi\;,\psi)\sqrt{1-\dot{\phi}^2/\dot{\phi}^2}+V(\phi\;,\psi)]$. The energy and density and pressure are $\rho=-p=K(\phi\;,\psi)\sqrt{1-\dot{\phi}^2/\dot{\phi}^2}+V(\phi\;,\psi)=const$ irrespective of the concrete form of $K(\phi\;,\psi)$ and $V(\phi\;,\psi)$. Actually, even if in the anisotropic but homogenous universe, the above quotient tachyonic scalars contribute a constant energy density and pressure.

However, we emphasize that in the anisotropic and inhomogeneous universe, the behavior of these scalar fields would be different from the Einstein cosmological constant. Therefore, a rigorous and exhaustive study of cosmological perturbations is needed in order to confirm different observational implications from cosmological constant in the large scale structure surveys. In the background of static and spherically symmetric spacetime, we have shown that the quotient scalar fields play the role of cosmological constant. How about for the background of stationary and axial symmetry spacetime? We conjecture that the quotient scalar fields would bring us with more than Kerr--de Sitter solution, they bring us with nontrivial Kerr solution dressed with scalar hairs.

Since the quotient scalar fields can play the role of cosmological constant in FRW universe, the cosmological constant can be abandoned. So the fine-tuning problem disappears. On the other hand, taking into account the interactions between the quotient scalar fields and matter or dark matter, the coincidence problem \cite{car:2001} can be alleviated.

\section{Acknowledgments}
This work is partially supported by the NSFC under grants 11633004 and 11773031.

\newcommand\arctanh[3]{~arctanh.{\bf ~#1}, #2~ (#3)}
\newcommand\ARNPS[3]{~Ann. Rev. Nucl. Part. Sci.{\bf ~#1}, #2~ (#3)}
\newcommand\AL[3]{~Astron. Lett.{\bf ~#1}, #2~ (#3)}
\newcommand\AP[3]{~Astropart. Phys.{\bf ~#1}, #2~ (#3)}
\newcommand\AJ[3]{~Astron. J.{\bf ~#1}, #2~(#3)}
\newcommand\GC[3]{~Grav. Cosmol.{\bf ~#1}, #2~(#3)}
\newcommand\APJ[3]{~Astrophys. J.{\bf ~#1}, #2~ (#3)}
\newcommand\APJL[3]{~Astrophys. J. Lett. {\bf ~#1}, L#2~(#3)}
\newcommand\APJS[3]{~Astrophys. J. Suppl. Ser.{\bf ~#1}, #2~(#3)}
\newcommand\JHEP[3]{~JHEP.{\bf ~#1}, #2~(#3)}
\newcommand\JMP[3]{~J. Math. Phys. {\bf ~#1}, #2~(#3)}
\newcommand\JCAP[3]{~JCAP {\bf ~#1}, #2~ (#3)}
\newcommand\LRR[3]{~Living Rev. Relativity. {\bf ~#1}, #2~ (#3)}
\newcommand\MNRAS[3]{~Mon. Not. R. Astron. Soc.{\bf ~#1}, #2~(#3)}
\newcommand\MNRASL[3]{~Mon. Not. R. Astron. Soc.{\bf ~#1}, L#2~(#3)}
\newcommand\NPB[3]{~Nucl. Phys. B{\bf ~#1}, #2~(#3)}
\newcommand\CMP[3]{~Comm. Math. Phys.{\bf ~#1}, #2~(#3)}
\newcommand\CQG[3]{~Class. Quantum Grav.{\bf ~#1}, #2~(#3)}
\newcommand\PLB[3]{~Phys. Lett. B{\bf ~#1}, #2~(#3)}
\newcommand\PRL[3]{~Phys. Rev. Lett.{\bf ~#1}, #2~(#3)}
\newcommand\PR[3]{~Phys. Rep.{\bf ~#1}, #2~(#3)}
\newcommand\PRd[3]{~Phys. Rev.{\bf ~#1}, #2~(#3)}
\newcommand\PRD[3]{~Phys. Rev. D{\bf ~#1}, #2~(#3)}
\newcommand\RMP[3]{~Rev. Mod. Phys.{\bf ~#1}, #2~(#3)}
\newcommand\SJNP[3]{~Sov. J. Nucl. Phys.{\bf ~#1}, #2~(#3)}
\newcommand\ZPC[3]{~Z. Phys. C{\bf ~#1}, #2~(#3)}
\newcommand\IJGMP[3]{~Int. J. Geom. Meth. Mod. Phys.{\bf ~#1}, #2~(#3)}
\newcommand\IJMPD[3]{~Int. J. Mod. Phys. D{\bf ~#1}, #2~(#3)}
\newcommand\IJMPA[3]{~Int. J. Mod. Phys. A{\bf ~#1}, #2~(#3)}
\newcommand\GRG[3]{~Gen. Rel. Grav.{\bf ~#1}, #2~(#3)}
\newcommand\EPJC[3]{~Eur. Phys. J. C{\bf ~#1}, #2~(#3)}
\newcommand\PRSLA[3]{~Proc. Roy. Soc. Lond. A {\bf ~#1}, #2~(#3)}
\newcommand\AHEP[3]{~Adv. High Energy Phys.{\bf ~#1}, #2~(#3)}
\newcommand\Pramana[3]{~Pramana.{\bf ~#1}, #2~(#3)}
\newcommand\PTP[3]{~Prog. Theor. Phys{\bf ~#1}, #2~(#3)}
\newcommand\APPS[3]{~Acta Phys. Polon. Supp.{\bf ~#1}, #2~(#3)}
\newcommand\ANP[3]{~Annals Phys.{\bf ~#1}, #2~(#3)}
\newcommand\RPP[3]{~Rept. Prog. Phys. {\bf ~#1}, #2~(#3)}
\newcommand\ZP[3]{~Z. Phys. {\bf ~#1}, #2~(#3)}
\newcommand\NCBS[3]{~Nuovo Cimento B Serie {\bf ~#1}, #2~(#3)}
\newcommand\AAP[3]{~Astron. Astrophys.{\bf ~#1}, #2~(#3)}
\newcommand\MPLA[3]{~Mod. Phys. Lett. A.{\bf ~#1}, #2~(#3)}


\begin{thebibliography}{99}
\bibitem{guth:1981} A. H. Guth, ``The Inflationary Universe: A Possible Solution to the Horizon and Flatness Problems,"
Phys. Rev. D 23 (1981) 347. doi:10.1103/PhysRevD.23.347


\bibitem{linde:1982} A. D. Linde, ``A New Inflationary Universe Scenario: A Possible Solution of the Horizon, Flatness,
Homogeneity, Isotropy and Primordial Monopole Problems," Phys. Lett. 108B, 389 (1982).
doi:10.1016/0370-2693(82)91219-9


\bibitem{ak:2018} Y. Akrami et al. [Planck Collaboration], ``Planck 2018 results. X. Constraints on inflation," arXiv:1807.06211 [astro-ph.CO].

\bibitem{perl:1998}S. Perlmutter et al. (Supernova Cosmology Project),
``Measurements of $\Omega$ and $\Lambda$ from 42 high redshift supernovae," Astrophys. J. 517, 565-586 (1999), arXiv:astro-ph/9812133.



\bibitem{adam:1998} Adam G. Riess et al. (Supernova Search Team), ``Observational evidence from supernovae for an accelerating
universe and a cosmological constant," Astron. J. 116, 1009-1038 (1998), arXiv:astro-ph/9805201.

\bibitem{fri:2008} J. A. Frieman, M. S. Turner, and D. Huterer, ``Dark Energy and the Accelerating Universe'', Annu. Rev.
Astron. Astrophys. 46, 385 (2008), arXiv:0803.0982 [astro-ph].



\bibitem{bart:2010}  M. Bartelmann, ``The Dark Universe'', Rev. Mod. Phys. 82, 331 (2010), arXiv:0906.5036 [astro-ph].

\bibitem{jer:2012} Jerome Martin, ``Everything You Always Wanted To
Know About The Cosmological Constant Problem (But
Were Afraid To Ask)," Comptes Rendus Physique 13,
566-665 (2012), arXiv:1205.3365 [astro-ph].

\bibitem{bur:2015}  C. P. Burgess, ``The Cosmological Constant Problem:
Why it's hard to get Dark Energy from Micro-physics," in
100e Ecole d'Ete de Physique: Post-Planck Cosmology
(2015) pp. 149-197, arXiv:1309.4133 [hep-th].


\bibitem{cop:2006} Edmund J. Copeland, M. Sami, and Shinji Tsujikawa,
``Dynamics of dark energy," Int. J. Mod. Phys. D 15, 1753-1936 (2006), arXiv:hep-th/0603057.


\bibitem{tim:2012} Timothy Clifton, Pedro G. Ferreira, Antonio Padilla, and Constantinos Skordis, ``Modified Gravity and Cosmology,"
Phys. Rept. 513, 1-189 (2012), arXiv:1106.2476 [astro-ph].

\bibitem{aus:2015} Austin Joyce, Bhuvnesh Jain, Justin Khoury, and Mark Trodden, ``Beyond the Cosmological Standard Model," Phys. Rept. 568, 1-98 (2015), arXiv:1407.0059 [astro-ph.CO].

\bibitem{Nojiri:2010wj}
S.~Nojiri and S.~D.~Odintsov, ``Unified cosmic history in modified gravity: from F(R) theory to Lorentz non-invariant models",
Phys. Rept. \textbf{505} (2011), 59-144
doi:10.1016/j.physrep.2011.04.001
[arXiv:1011.0544 [gr-qc]].

\bibitem{Bamba:2012cp}
K.~Bamba, S.~Capozziello, S.~Nojiri and S.~D.~Odintsov,
``Dark energy cosmology: the equivalent description via different theoretical models and cosmography tests,''
Astrophys. Space Sci. \textbf{342} (2012), 155-228
doi:10.1007/s10509-012-1181-8
[arXiv:1205.3421 [gr-qc]].



\bibitem{Nojiri:2017ncd}
S.~Nojiri, S.~D.~Odintsov and V.~K.~Oikonomou,
``Modified Gravity Theories on a Nutshell: Inflation, Bounce and Late-time Evolution,''
Phys. Rept. \textbf{692} (2017), 1-104
doi:10.1016/j.physrep.2017.06.001
[arXiv:1705.11098 [gr-qc]].



\bibitem{phi:2016}Philip Bull, Yashar Akrami, et al., ``Beyond $\Lambda$CDM: Problems, solutions, and the road ahead," Phys. Dark Univ.
12, 56-99 (2016), arXiv:1512.05356 [astro-ph].

\bibitem{ban:2021} Aritra Banerjee, Haiying Cai, et al., ``Hubble Sinks In The Low-Redshift Swampland", Phys. Rev. D 103, 081305 (2021), arXiv:2006.00244 [astro-ph]
\bibitem{bau:0907} D. Baumann, ``TASI Lectures on Inflation", [arXiv:0907.5424].

\bibitem{langlois:2008} D. Langlois and S. Renaux-Petel, ``Perturbations in generalized multi-field inflation'',\JCAP{0804}{017}{2008}. [arXiv: hep-th/0801.1085].
\bibitem{linde:1983} A. D. Linde, ``Chaotic inflation'', Phys. Lett. B 129, 177 (1983). doi.org/10.1016/0370-2693(83)90837-7

\bibitem{akr:2020} Y. Akrami et al., ``Planck 2018 results. X. Constraints on inflation'', \AAP{641}{A10}{2020}, arXiv:1807.06211 [astro-ph].


\bibitem{ade:2018} P. A. R. Ade et al., ``BICEP2/Keck Array x: Constraints on Primordial Gravitational Waves using Planck, WMAP, and New BICEP2/Keck Observations through the 2015 Season'', \PRL{121}{221301}{2018}, arXiv:1810.05216 [astro-ph].


\bibitem{car:2001}S. M. Carroll, ``The Cosmological Constant'', Living Rev. Relativity 4, 1 (2001), arXiv:astro-ph/0004075 [astro-ph]





\end{thebibliography}
\end{document}